\def\year{2020}\relax
\documentclass[letterpaper]{article} 
\newcommand{\projectName}{\textit{Kanji Workbook}}

\usepackage{aaai20}  
\usepackage{times}  
\usepackage{helvet} 
\usepackage{courier}  
\usepackage[hyphens]{url}  
\usepackage{graphicx} 
\usepackage{graphicx}  
\usepackage{booktabs}
\usepackage{multirow}
\usepackage[super]{nth}

\title{
    Kanji Workbook: A Writing-Based Intelligent Tutoring System for Learning
    Proper Japanese Kanji Writing Technique with Instructor-Emulated Assessment
}

\author{
    Paul Taele, Jung In Koh, Tracy Hammond \\
    Sketch Recognition Lab \\
    Department of Computer Science \& Engineering \\
    Texas A\&M University \\
    \{ptaele, jungin, hammond\}@tamu.edu
}

\begin{document}

\maketitle

\begin{abstract}
Kanji script writing is a skill that is often introduced to novice Japanese foreign language students for achieving Japanese writing mastery, but often poses difficulties to students with primarily English fluency due to their its vast differences with written English.
Instructors often introduce various pedagogical methods---such as visual structure and written techniques---to assist students in kanji study, but may lack availability providing direct feedback on students' writing outside of class.
Current educational applications are also limited due to lacking richer instructor-emulated feedback.
We introduce \projectName{}, a writing-based intelligent tutoring system for students to receive intelligent assessment that emulates human instructor feedback.
Our interface not only leverages students' computing devices for allowing them to learn, practice, and review the writing of prompted characters from their course's kanji script lessons, but also provides a diverse set of writing assessment metrics---derived from instructor interviews and classroom observation insights---through intelligent scoring and visual animations.
We deployed our interface onto novice- and intermediate-level university courses over an entire academic year, and observed that interface users on average achieved higher course grades than their peers and also reacted positively to our interface's various features.
\end{abstract}

\section{Introduction}
For students studying Japanese as a foreign language, one of the main components for achieving mastery of the language is fluency in reading and writing the language.
Unlike in English and its sole script of the Latin alphabet, the Japanese language heavily utilizes three separate scripts, including the kanji script and its characters that differ greatly from alphabet letters in English.
Not only do kanji script characters have characteristics that challenge language students such as their large number, diverse structure, and tendency to have visual similarity among each other~\cite{Taele:2009:IAAI:Hashigo}, but is one of several factors that makes the Japanese language one of the most challenging languages for native English users to learn~\cite{NSA:2017:Misc:LanguageDifficulty}.

In order to address these challenges, language instructors often employ various pedagogical approaches through stroke writing techniques for enabling students to improve their mastery in kanji script characters~\cite{Ye:2011:Dissertation:TeachingCharacters}, since such techniques are important in Japanese language writing~\cite{Takezaki:2008:Book:JapaneseCalligraphy}.
However, for conventional classroom settings, instructors may not have the resources to provide direct assistance on assessing students' character writing performance with large classes or outside of class~\cite{Taele:2009:IAAI:Hashigo}, and existing educational applications for character writing practice remain constrained on the depth of assessment that emulates those of human language instructors.

In this work, we propose \projectName{}, a writing-based intelligent tutoring for supporting novice Japanese foreign language students in their study and practice of kanji script character writing.
Our work not only allows students to employ writing interactions for their study of the characters, but also provides a variety of assessment metrics that provide more granular feedback similarly to human instructors.
We designed our interface based on qualitative insights from language instructors, and deployed the interface onto novice university courses in Japanese as a foreign language.
From the classroom and interaction studies for evaluating our interface, we observed that students benefited with improved academic performance on their character writing capabilities, and discovered that students found the interface to be intuitive and the learning to be rewarding.

\section{Related Work}
As a writing-based intelligent tutoring system, \projectName{} relates to other educational applications in terms of factors such as pedagogical goals, interface features, and recognition techniques.
These applications include Mechanix~\cite{Valentine:2012:IAAI:Mechanix} for civil engineering, iCanDraw~\cite{Dixon:2010:CHI:iCanDraw} for art drawing, MathPad2~\cite{LaViola:2004:SIGGRAPH:MathPad2} for mathematical formulas, and Maestoso~\cite{Taele:2015:IAAI:Maestoso} for music composition.
While such applications share similar themes to \projectName{}, they provide educational assistance for domains that are not directly related to foreign language writing.

Since \projectName{} also provides recognition capabilities for kanji script character writing, works on automated recognition of East Asian languages that use characters such as Japanese and Chinese also relate to \projectName{}.
Works from this research area that have focused on recognition techniques for character scripts have continued to exist for the past several decades, and leverage machine learning approaches ranging from neural networks to hidden Markov models~\cite{Zhang:2017:PR:ChineseRecognition}.
Furthermore, systems that incorporate or advance the approaches from these works have become familiar in commercial tools that are available to the public~\cite{Pittman:2007:Computer:TabletRecognition}.
However, systems that incorporate these approaches focus more on optimizing the classification rates of casual handwriting of fluent writers, as opposed to providing more human-like assessment of novice handwriting of foreign language students.

Writing-based systems that both provide the intelligence as educational applications and focus on character scripts such as the Japanese language's kanji script have seen exploration since the 1990s~\cite{Ma:2009:ICDAR:ChineseHMM}, and many of these works focus on assessing the visual correctness of character writing.
Other systems have considered the technical production of the character's strokes~\cite{An:2011:ICEEE:ChineseStrokes,Taele:2009:IAAI:Hashigo,Taele:2010:JVLC:LAMPS,Chu:2018:GI:ChineseCharacters}, but limitations still linger.
Specifically, the depth of assessment on students' character writing tends to focus only on a limited set of popular pedagogical approaches such as stroke order and stroke direction, while other types of assessments---such as for visual structure and writing precision---are little considered.
As a result, \projectName{} aims to provide a writing-based intelligent tutoring system with richer assessment of students' character writing performance that better emulate those provided in human language instructors.

\section{Qualitative Study}
Prior to developing our interface, we first conducted a qualitative study---consisting of instructor interviews and classroom observations---to discover real-world classroom best practices for character writing instruction.
These practices would eventually be qualitatively analyzed into a list of insights that would lead to the design of the user interface and assessment system.
In the study, we interviewed four instructors and observed two classrooms, all from different university-level foreign language programs that offered Chinese character writing instruction.

The materials (i.e., transcripts and notes) produced from the semi-structured interviews and classroom observations were qualitatively analyzed into encoded data units~\cite{Glaser:1965:SP:QualitativeAnalysis}, where we define our units as concepts related to character writing instruction and also emphasized, repeated, or deemed important from the materials.
From our initial selection of 269 data units, we applied a constant comparative method~\cite{Glaser:1965:SP:QualitativeAnalysis} that organized these data units into initially 54 data codes, then seven categories, and finally three emergent themes.
Two of the emergent themes---\textit{in-class instruction} and \textit{learning activities \& resources}---led to insights that defined the interaction features of our user interface.
The remaining emergent theme---\textit{class challenges}---yielded valuable insights on a variety of students' common character writing mistakes that instructors considered important to assess for correctness.
This particular insight led to the development of ten assessment metrics that was used in our interface and is further elaborated in the following section. 

\section{Assessment System}
The assessment system provides the automated recognition of students' kanji script character writing performance through our proposed collection of assessment metrics.
Prior works have incorporated a small subset of the interface's assessment metrics for students' character writing such as stroke order and direction~\cite{Taele:2009:IAAI:Hashigo,Taele:2010:JVLC:LAMPS}.
One of the novel contributions in our work was further expanding these existing assessments to a total of ten metrics, based on the insights on our earlier qualitative study that confirmed and expanded upon the few assessment metrics employed of those past works.
Furthermore, these ten assessment metrics measured the students' character writing performance relative to model templates from an expert instructors' collection of character writing data.
The following goes into greater details of the AI techniques that were employed in our assessment system.

\subsection{Character Writing Data Model Templates}
In order to assess the students' character writing performance, their writing data was compared to model templates from an expert's character writing data.
Specifically, the data was collected from one of the Japanese foreign language instructors who helped consult with the development of the interface.
The expert was tasked with providing character writing data---consisting of the 448 characters that were used in the interface---as model templates for our assessment system.
The expert was instructed to write the characters as models for comparison with students' character writing data through a specialized data collection application.
The data format of the model character writing was stored in JSON format, and consisted of the following relevant components similarly utilized in prior sketch recognition systems such as LADDER~\cite{Hammond:2007:CG:LADDER}:

\begin{itemize}
    \item \textbf{Point.} The writing canvas samples points from the user's input contact on the interface's writing canvas.
    Each point collected is stored as an object with an x- and y-coordinate value as spatial data, and with a timestamp value as temporal data.
    \item \textbf{Stroke.} For each collection of points that was sampled from the user's start of writing (e.g., touching the screen) to end of writing (e.g., lifting off the screen), the interface groups them together as a stroke.
    \item \textbf{Sketch.} For each collection of strokes that was collected from an empty writing canvas to submission for assessment, the interface groups them together as a sketch.
    \item \textbf{Metadata.} Additional information is stored in the data as metadata for supplemental interface purposes.
    These metadata attributes include the character's identification label for retrieval purposes, and its original canvas dimensions for scaling to the dimensions of the interface's writing canvas.
\end{itemize}

Once the expert's character writing data was collected as model templates for our assessment system, this collection was then pre-processed through transformation algorithms.
This pre-processing step was performed to achieve faster assessment by normalizing the model template into a consistent size, so that only the students' character writing data---which can be input from devices of multiple screen sizes (e.g., smartphones, tablets)---during the assessment process.
For the normalization steps of the expert data, we leveraged the following transformation algorithms that were similar utilized in stroke-based template-matching recognizers~\cite{Wobbrock:2007:UIST:Dollar}:

\begin{itemize}
    \item \textbf{Resampling.} This transformation rearranges the template's stroke points into 64 equally-spaced points.
    \item \textbf{Scaling.} This transformation resizes the template's bounding box to a size of 250 pixels.
    \item \textbf{Translating.} This transformation shifts the template's bounding box to the origin in Cartesian system-based coordinates.
\end{itemize}

\subsection{Generalized Assessment Recognition Process}
In order for our interface to perform instructor-emulated assessments on the students' character writing input, the assessment system follows a series of automated processing and recognition steps.
The following list details a high-level summary description of each step for all the assessment metrics:

\begin{enumerate}
    \item \textbf{Retrieve Model Templates.} Once the user starts the interface, perform a one-time retrieval of the model templates from the server.
    \item \textbf{Record Input Data.} After the user writes each character and submits it for assessment, record the user's current character writing data into local data structures.
    \item \textbf{Normalize Input Data.} Run the same transformation algorithms that were used on the model templates on the user's input data.
    \item \textbf{Assess Input Data.} Perform each of the ten assessment metrics on the input data.
    \item \textbf{Calculate Input Data's Score.} Based on each assessment metric's calculated results, calculate the input data's performance score.
\end{enumerate}

More specialized steps are required for the ten assessment metrics.
Due to these metrics sharing similarity in both assessment concept and recognition process with other metrics, they have been grouped together into the following types: structure, technique, and precision.
The following sections provide further details about the three assessment types.

\subsection{Structure Assessments}
\begin{figure}[hbtp]
    \centering
    \includegraphics[height=2cm]{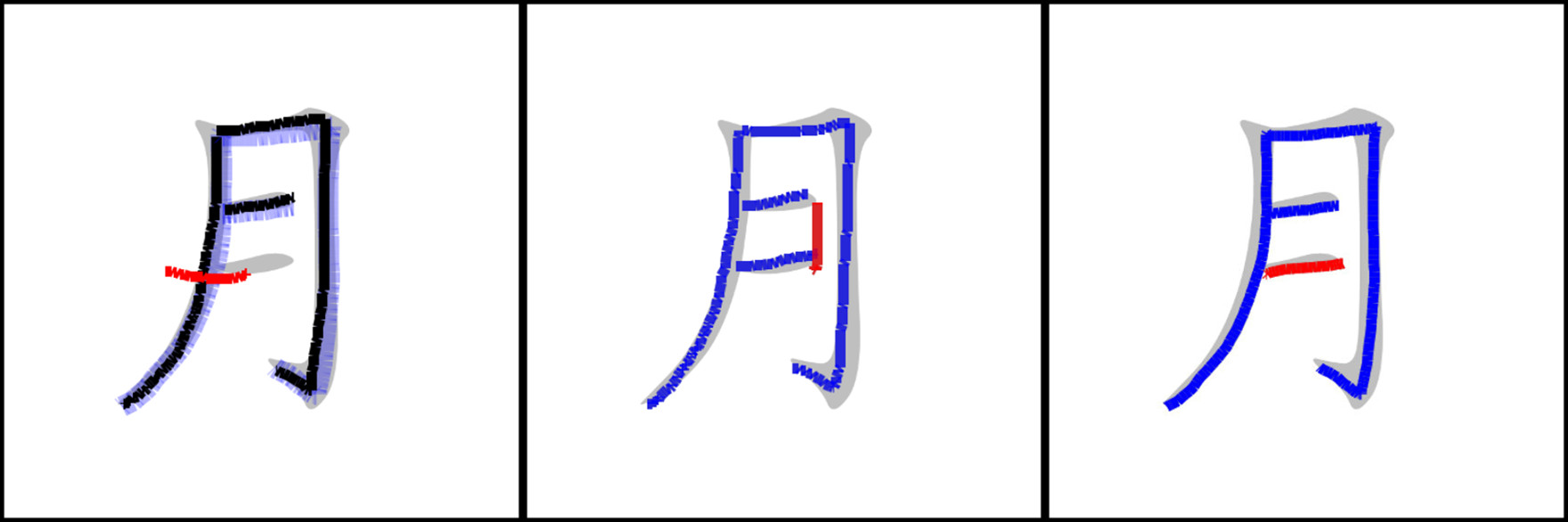}
    \caption{Structure assessment visualizations for: stroke match (left), stroke valid (middle), and stroke exist (right).}
    \label{figure:assessment:structure}
\end{figure}

The first set of assessment metrics focuses on the correctness of students' visual structure in their character writing.
These assessments are important to instructors for better informing them on whether a students' character writing is visually correct.
Instructors conventionally assess students on their visual structure by reviewing the students' input strokes on several factors such as whether they exist and are visually similar enough.
However, instructors are challenged in assessing students' visual structure such as determining lacking and extraneous strokes and whether a stroke was unambiguously written together or not.
Our assessment system accommodates instructors by providing metrics for the following assessments (Figure~\ref{figure:assessment:structure}): \textit{stroke match}, \textit{stroke valid}, and \textit{stroke exist}.

An important step for assessing the visual structure correctness of students' character writing is first finding the existing corresponding pairs between the students' input strokes to the expert's model strokes.
This step is achieved by leveraging a strategy that is commonly employed by stroke-based template matching algorithms for classifying different classes of pen and touch gestures.
Specifically, we leverage the Hausdorff distance metric for calculating the distance error between the user inputs and the model templates' corresponding points from relevant template matching algorithms such as the \$P gesture classifier~\cite{Vatavu:2012:ICMI:PDollar}.
However, instead of automatically classifying user's input gestures to labels of most likely corresponding model templates with the smallest distance error, we adapt this strategy to automatically classify students' strokes to indices of most likely corresponding expert's model strokes.

\begin{itemize}
    \item \textbf{Stroke Match.} This metric assesses how many correct matches exist between the user's input strokes and the expert's model strokes.
    \item \textbf{Stroke Valid.} This metric assesses how many model strokes correctly match to the user's corresponding input strokes, and also assists instructors in assessing whether a student has missing correct strokes.
    \item \textbf{Stroke Exist.} This metric assesses how many input strokes correctly match to the expert's corresponding model strokes, and also assists instructors in assessing whether a student has extraneous strokes.
\end{itemize}

\subsection{Technique Assessments}
\begin{figure}[hbtp]
    \centering
    \includegraphics[height=2cm]{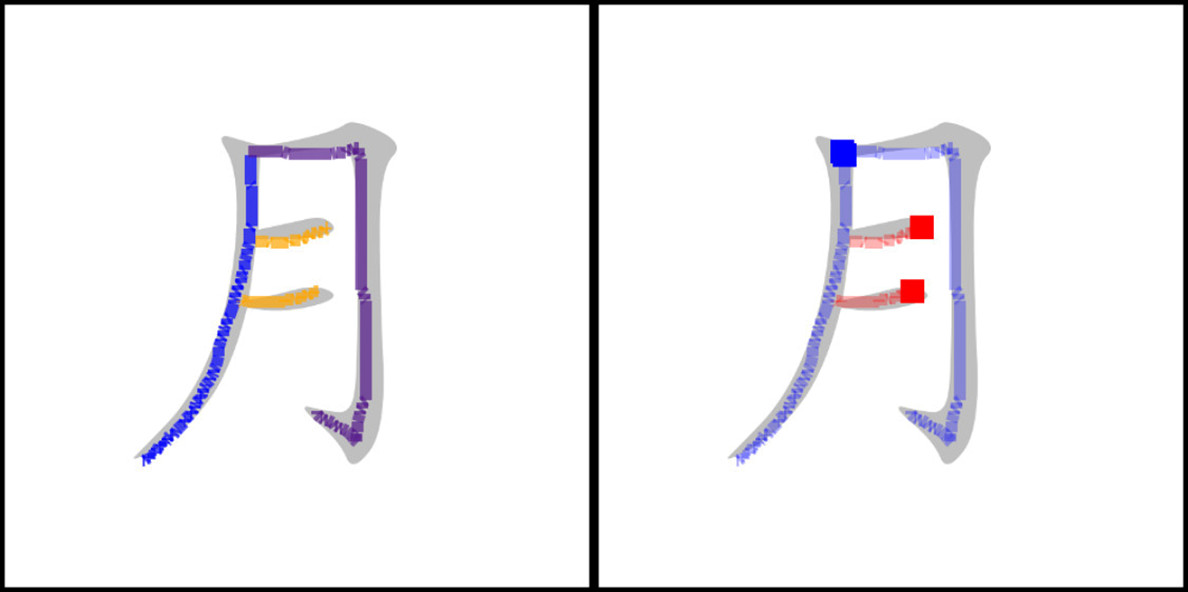}
    \caption{Technique assessment visualizations for: stroke order (left) and stroke direction (right).}
    \label{figure:assessment:technique}
\end{figure}

The second set of assessment metrics focuses on the correctness of students' written technique in their character writing.
These assessments are important to instructors for better informing them on whether a students' character writing is technically correct.
Instructors conventionally assess students on their visual structure by physically observing their character writing in class.
However, instructors are challenged in assessing students' visual structure due to time constraints of reviewing each students' character writing or less optimal solutions of requiring students to physically label their strokes in homework assignments.
Our assessment system accommodates instructors by providing metrics for the following assessments (Figure~\ref{figure:assessment:technique}): \textit{stroke order} and \textit{stroke direction}.

An important step for assessing the written technique correctness of students' character writing is first finding the existing correspondences of strokes and stroke endpoints between the students' input strokes to the expert's model strokes.
This step helps determine whether the student wrote their strokes temporally correct in both order and direction.
However, finding the correspondences of the endpoints between the user's input strokes and the expert's model strokes is non-trivial for assessing stroke direction, since there may be cases where both endpoints of a user's stroke may have similar distances to an endpoint of the corresponding expert stroke (e.g., small strokes, long strokes with close endpoints).
As a result, we leverage the Euclidean distance metric for calculating the distance error between the user's input stroke and the corresponding expert's model stroke from relevant template matching algorithms such as the \$1 gesture classifier~\cite{Vatavu:2012:ICMI:PDollar}.
This strategy differs from similar prior systems that also assess stroke direction~\cite{Taele:2009:IAAI:Hashigo} since we calculate the entire distance error to more accurately classify the correct corresponding endpoints.
From this strategy, we summarize the AI techniques employed by our technique assessment metrics:

\begin{itemize}
\item \textbf{Stroke Order.} This metric assesses how many of the user's input strokes are written in temporal correct order.
\item \textbf{Stroke Direction.} This metric assesses how many input of the user's input strokes are written in temporal correct direction.
\end{itemize}

\subsection{Precision Assessments}
\begin{figure}[hbtp]
    \centering
    \includegraphics[height=2cm]{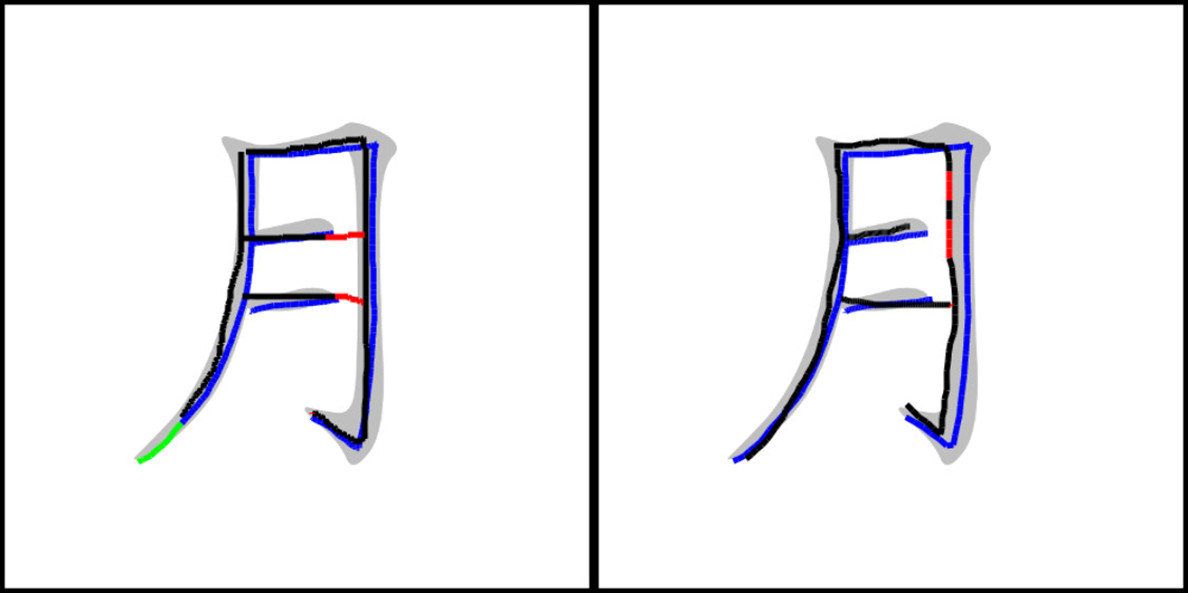}
    \caption{Precision assessment visualizations for: stroke length (left) and stroke closeness (right).}
    \label{figure:assessment:precision}
\end{figure}

The third set of assessment metrics focuses on the correctness of students' precision performance in their character writing.
These assessments are important to instructors for better informing them on whether a student is seamlessly writing the characters.
Instructors conventionally assess students on their precision performance through direct observation, but becomes a challenge for consistent assessment physical presence for larger class sizes.
Our assessment system accommodates instructors by providing metrics for the following assessments (Figure~\ref{figure:assessment:precision}): \textit{stroke edit}, \textit{stroke speed}, \textit{stroke length}, \textit{stroke closeness}, and \textit{symbol speed}.

An important step for assessing the precision performance of students' character writing is determining the appropriate thresholds of the different assessment metrics similar to the assessment feedback of an instructor.
To achieve this type of assessment, we employed a similar strategy found in sketch recognition systems such as~\cite{Hammond:2007:CG:LADDER} and its follow-up works that assigns pre-defined thresholds for feature-based classification of sketches.
We adapt this strategy of our own pre-defined thresholds that were empirically derived from the expert model templates to develop features that classify the assessment level of students' character writing.
From this strategy, we summarize the AI techniques employed by our technique assessment metrics:

\begin{itemize}
    \item \textbf{Stroke Edit.} This metric assesses how many times the student edits their input strokes during their character writing session.
    \item \textbf{Stroke Length and Closeness.} These metrics assess the stroke length and spatial distances, respectively, between each matched pair of input and model strokes.
    \item \textbf{Stroke and Symbol Speed.} These metrics assess the differences in relative writing time duration between the input and models' strokes and characters, respectively.
\end{itemize}

\section{User Interface}
For \textit{Kanji Workbook}, we implemented a user interface that takes inspirational design cues from systems for other domains~\cite{Dixon:2010:CHI:iCanDraw,LaViola:2004:SIGGRAPH:MathPad2,Valentine:2012:IAAI:Mechanix}, and also adapted similar visual and interaction cues to other related systems~\cite{Taele:2009:IAAI:Hashigo,Sloniger:GI:TensaiGame:2018}.
We also incorporated novel components and features in our user interface that is elaborated in the following.

\begin{figure}[hbtp]
    \centering
    \includegraphics[width=0.6\columnwidth]{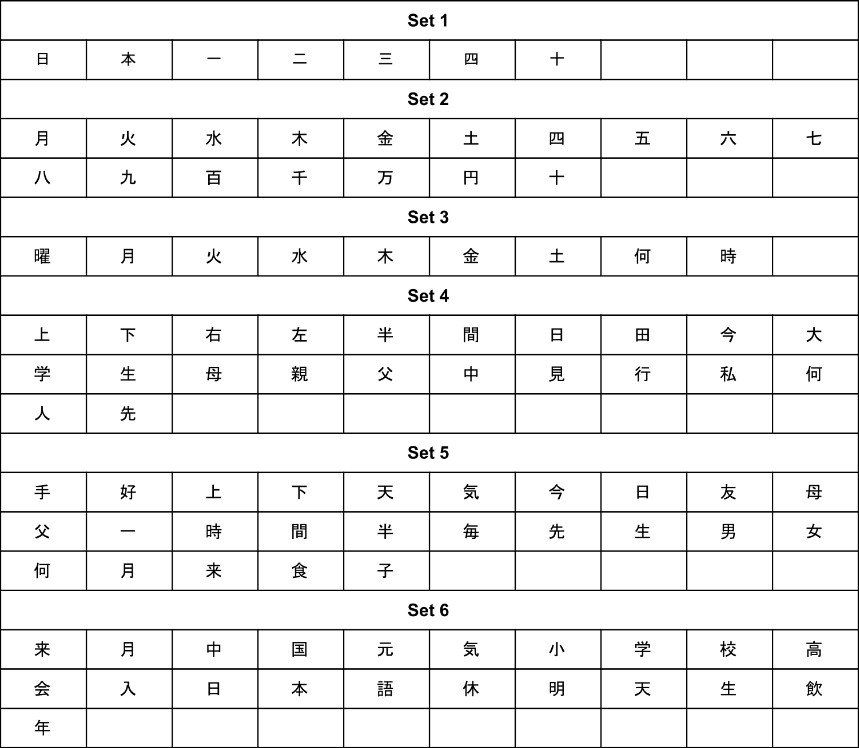}
    \caption{Subset of characters in \projectName{}.}
    \label{figure:textbook:character:subset}
\end{figure}

\subsection{Lesson Characters}
The kanji script characters that were adapted in the interface were derived from the lesson chapters in the Genki textbook series~\cite{Banno:2011:Book:GenkiI,Banno:2011:Book:GenkiII} (Figure~\ref{figure:textbook:character:subset}).
This textbook was utilized since it was the required textbook for the language courses that deployed the interface (see \textit{Deployment Results}).
The interface corresponds to the characters used in the chapters from the Genki textbooks for a total of 23 selectable lessons, which consisted of 448 characters combined.

\begin{figure}[hbtp]
    \centering
    \includegraphics[width=0.9\columnwidth]{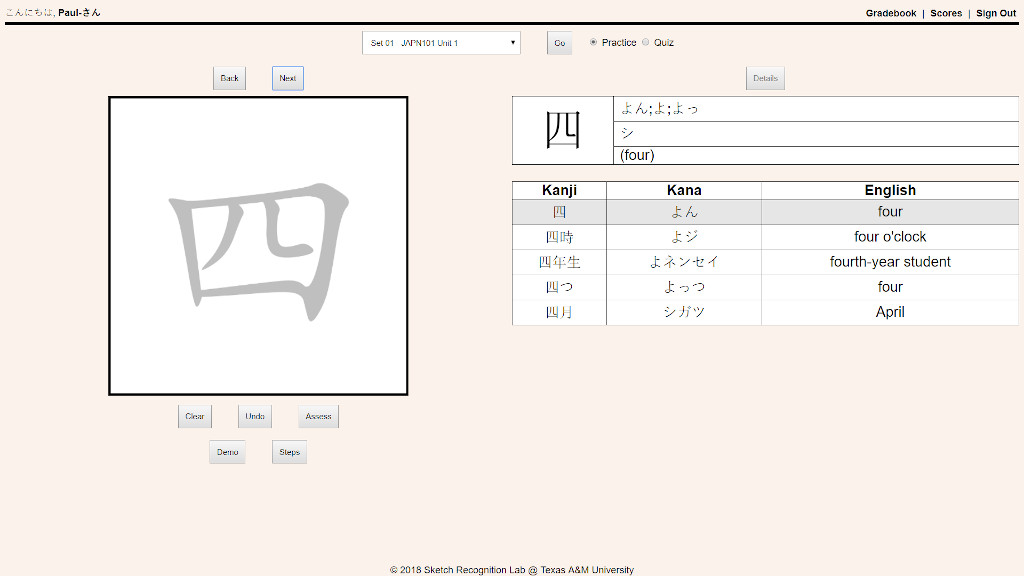}
    \caption{Practice mode view.}
    \label{figure:interface:practice:mode:view}
\end{figure}

\begin{figure}[hbtp]
    \centering
    \includegraphics[width=0.9\columnwidth]{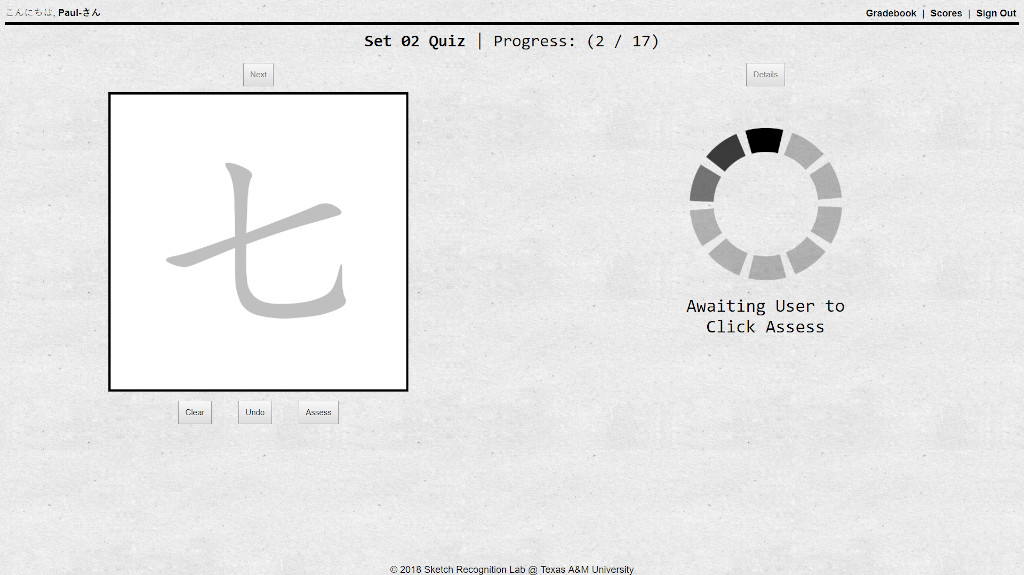}
    \caption{Quiz mode view.}
    \label{figure:interface:quiz:mode:view}
\end{figure}

\subsection{Writing Area Space}
In some views of the interface, there is an area for users to provide their kanji script character writing on the left half portion of the interface (e.g., Figure~\ref{figure:interface:practice:mode:view}).
This writing space area is devoted to two smaller areas worth highlighting.

\subsubsection{Writing Canvas}
The first smaller area consists of the writing canvas, which is an interaction space for users to write the characters using stylus, touch, or mouse.
The writing canvas also displays a semi-transparent visualization of the current character in the lesson for the user to trace over, and also displays animations and feedback strokes that overlay the user's own character writing.

\subsubsection{Interaction Buttons}
The second smaller area consists of interaction buttons that surround the writing canvas. Conventional buttons include navigation (i.e., \textit{Back}, \textit{Next}), editing (i.e., \textit{Clear}, \textit{Undo}), and submit (i.e., \textit{Assess}). Two addition buttons display informative animations: \textit{Demo}, for showing the full character writing animation; and \textit{Steps}, for a similar stroke-by-stroke animation.

\subsection{Character Information Space}
While the user is practicing character writing, the right half portion of the interface displays supplemental information about the character (Figure~\ref{figure:interface:practice:mode:view}).

\begin{itemize}
    \item \textbf{Character.} A text visualization of the character.
    \item \textbf{Pronunciations/Translations.} The Japanese pronunciations and English translations of the character.
    \item \textbf{Vocabulary.} The vocabulary words from the Genki textbooks that use the character, and their corresponding Japanese pronunciations and English translations. The highlighted vocabulary words indicate words that are tested in the textbooks' supplemental paper workbooks.
\end{itemize}

\subsection{Assessment Area Space}
\begin{figure}[hbtp]
    \centering
    \includegraphics[width=0.9\columnwidth]{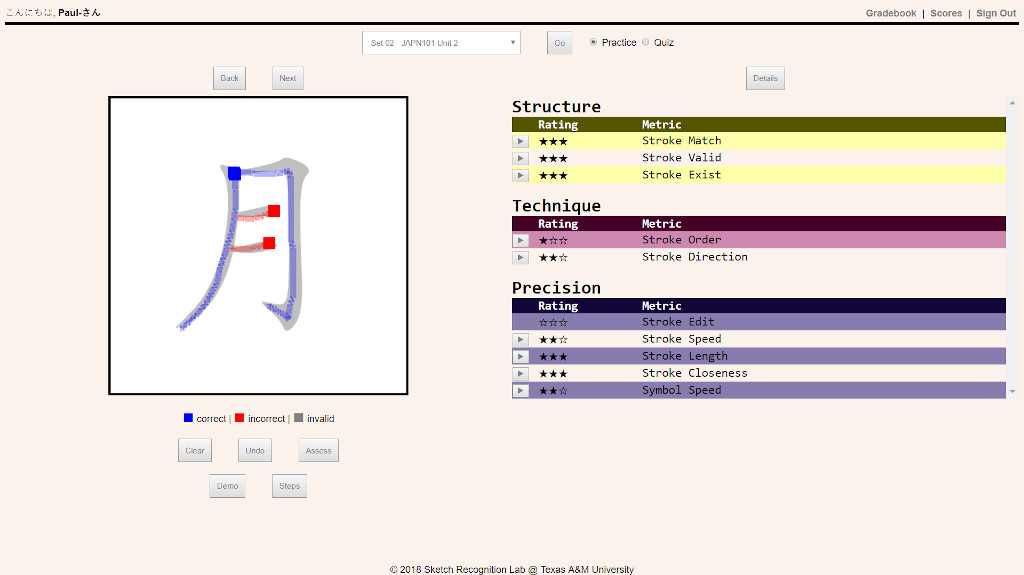}
    \caption{Character results view.}
    \label{figure:interface:character:results:view}
\end{figure}

Once a user submits a character for assessment, the right half portion of the interface displays assessment information on the user's character writing performance (Figure~\ref{figure:interface:character:results:view}).
More specifically, this space shows each of the assessment metrics and their corresponding animation buttons and scoring scores.
Details of these metrics are described in the \textit{Assessment System} section.

\begin{itemize}
    \item \textbf{Animation Buttons} These buttons displays an animation that relates to their corresponding assessment metric.
    \item \textbf{Scoring Stars} This area shows a visualization of three stars for providing a three-point score of the user's character writing performance, where the number of displayed stars indicate the level of correctness for that particular assessment metric.
    \item \textbf{Color Key} This visualization is located in the writing area space, and is dynamically shown for certain animations to inform users of the purpose of the different color strokes that are displayed in the animation.
\end{itemize}

\subsection{Lesson Modes}
Two modes are available in the interface for users to accommodate their lesson studies, in order to emulate their existing character writing practice and quiz activities .

\begin{figure}[hbtp]
    \centering
    \includegraphics[width=0.9\columnwidth]{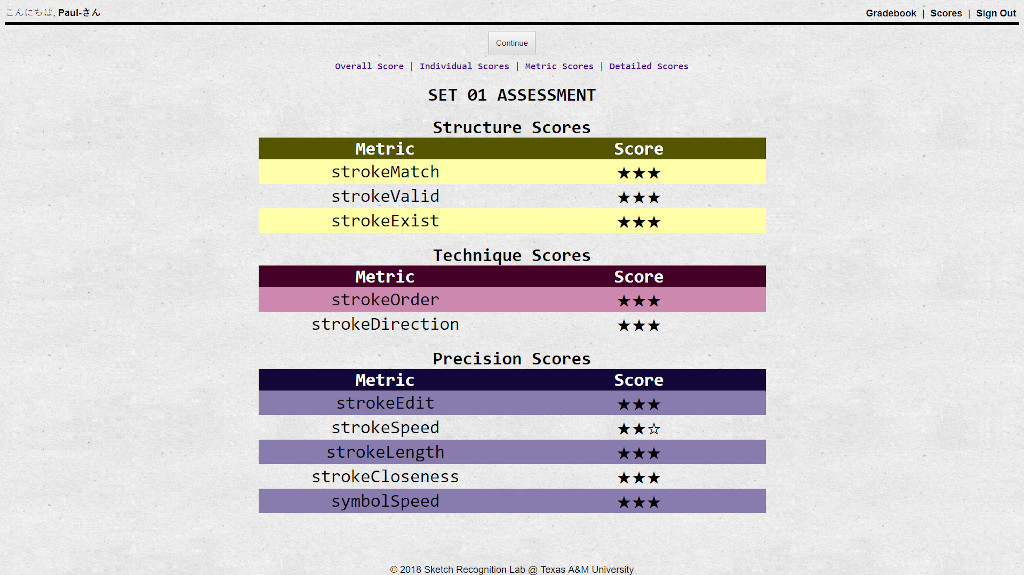}
    \caption{Quiz results view.}
    \label{figure:interface:quiz:results:view}
\end{figure}

\begin{itemize}
    \item \textbf{Practice Mode.} The first lesson mode is the practice mode, which allows users to pursue character writing study in a workbook writing-like activity (Figure~\ref{figure:interface:practice:mode:view}).
    In this mode, users are able to progress forwards and backwards for characters multiple times in the lesson, and are also able to view supplementary character information during their writing.
    \item \textbf{Quiz Mode.} The second lesson mode is the quiz mode, which allows users to pursue character writing study in a quiz taking-like activity (Figure~\ref{figure:interface:quiz:mode:view}).
    In this mode, users can only progress forward for characters once in the lesson, are not shown supplementary character information during their writing.
    Furthermore, they receive a comprehensive assessment of their lesson-wide character writing performance at the end of the lesson (Figure~\ref{figure:interface:quiz:results:view}).
\end{itemize}

\section{Evaluation}

\subsection{Classroom Study}
Following the interface's completed development, the interface was deployed onto \nth{1}- and \nth{2}-year Japanese foreign language courses at Texas A\&M University's Department of International Studies.
The study was conducted in collaboration with two instructors teaching these \nth{1}- and \nth{2}-year courses, respectively, where enrolled students were observed over a time period spanning nine consecutive months (i.e., during the Fall 2018 and Spring 2019 academic semesters).
A total of 94 students---29 females---from these courses took part in the classroom study.
These participating students were partitioned into case and control groups, where the case group were tasked with using the interface in their kanji script character writing study for their enrolled semester, and the control group was tasked with using conventional study habits with paper-based materials and optional online resources,  (Table~\ref{table:classroom:study:participant:count}).

\begin{table}[hbtp]
\footnotesize
\centering
\caption{Total number of classroom study participants.}
\label{table:classroom:study:participant:count}
\begin{tabular}{|c|c|c|c|}
\hline
\textbf{LANGUAGE YEAR} & \textbf{CONTROL} & \textbf{CASE} & \textbf{TOTAL} \\ \hline
1st Year only          & 34               & 16            & 50             \\ \hline
2nd Year only          & 16               & 28            & 44             \\ \hline
Total                  & 50               & 44            & 94             \\ \hline
\end{tabular}
\end{table}

During the study, students from both case and control groups were introduced to kanji script characters for study, as normally scheduled by their instructor in the curriculum.
Over the span of their enrolled semesters, students were assigned five in-class tests with strong emphasis on vocabulary and sentence writing from the latest introduced characters.
Students in the control group were instructed to utilize paper workbook assignments and any optional educational resources for learning the characters.
Those in the case group were assigned to log into the interface and encouraged to practice their character writing on the interface's practice mode sessions for learning the characters, in addition to their required paper workbook assignments.

\begin{figure}[hbtp]
    \centering
    \includegraphics{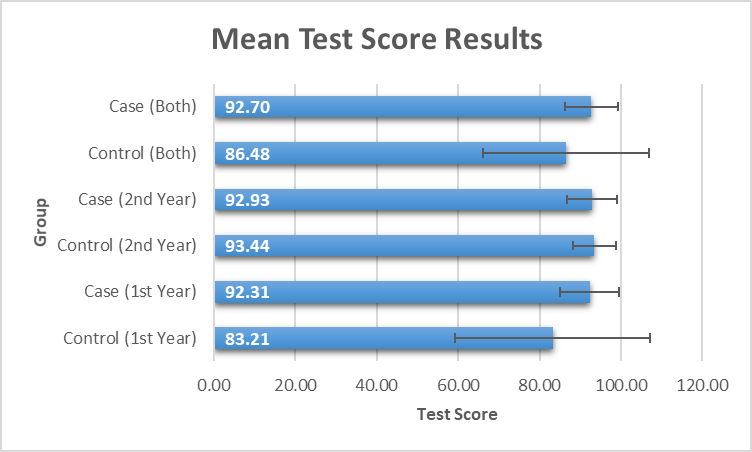}
    \caption{Mean test score results.}
    \label{figure:mean:test:score:results}
\end{figure}

At the end of the deployment period, the pre-adjusted averaged test scores---on a 100-point scale from the five semester tests---were shared by the collaborating instructors of the \nth{1}- and \nth{2}-year courses (Figure~\ref{figure:mean:test:score:results}).
From these scores:

\begin{itemize}
    \item the \nth{1}-year case group scored higher on average than the \nth{1}-year control group by 9.10 points
    \item the \nth{2}-year case group differed on average from the \nth{2}-year control group by 0.51 points
    \item the entire case group scored higher on average than the entire control group by 6.22 points
\end{itemize}

One noticeable observation was that the \nth{1}-year control group had a large standard deviation, due to the wide range of scores between low- and high-performing students in the \nth{1}-year control group.
In order to determine whether this observation affected whether the entire case group's higher average scores were statistically significant, a t-test was conducted between the scores of the entire case group and the entire control group.
From the two-tailed t-test, the calculated p-value was p=.046$<$.05, which demonstrated that the test scores from the entire case group was still considered statistically significant (Table~\ref{table:test:score:results:statistics}) despite the observation.
When the two-tailed t-test was similarly performed for the test scores between the test scores of the \nth{1}-year control group and the \nth{1}-year case group, the calculated p-value of p=.048$<$.05 also demonstrated statistical significance for the \nth{1}-year case group's higher average scores compared to the \nth{1}-year control group.

\begin{table}[!hbtp]
\footnotesize
\centering
\caption{Test score results statistics for test scores between the entire case group and the entire control group.}
\label{table:test:score:results:statistics}
\begin{tabular}{@{}lll@{}}
\toprule
\textbf{Statistics}          & \textbf{Control} & \textbf{Case} \\ \midrule
Mean                         & 86.48            & 92.70         \\
Variance                     & 416.7            & 42.26         \\
Observations                 & 50               & 44            \\
P(T\textless{}=t) two-tail   & 0.046            &               \\
\bottomrule
\end{tabular}
\end{table}

\subsection{Interaction Study}
An interaction study was conducted that asked students to provide Likert-scale and freeform responses---with constructive feedback on the interface's features of buttons, ratings, and animations---in terms of intuitiveness and usefulness.
For the Likert-scale responses, most students agreed or strongly agreed on the intuitiveness and usefulness of these three features.
For the freemform, students shared positive feedback on the interface for providing a self-improving and rewarding experience while practicing their character writing, and also shared that the interface overall was intuitive and that the feedback and animations were useful.
Of the three interface features, the ratings feature---which consisted of the three-star scoring system---received relatively lower Likert-scale agreement scores in terms of intuitiveness.
That is, less students provided \textit{Strongly Agree} responses for the ratings' intuitiveness, though the majority of students provided \textit{Strongly Agree} responses for all the ratings in terms of usefulness.
The freeform responses on the ratings' lower agreement scores for intuitiveness revealed that students either generally felt that the ratings were overly strict, or that ratings for specific assessment metrics (i.e., stroke speed, order, closeness, and length) were less intuitive.

\subsection{Recognized Improper Characters}
\begin{figure}[hbtp]
    \centering
    \includegraphics[width=0.7\columnwidth]{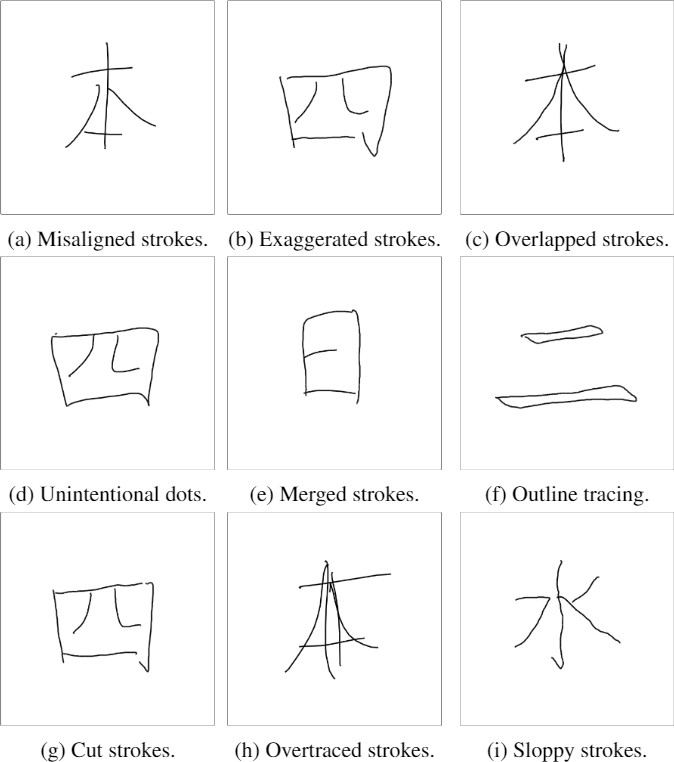}
    \caption{Examples of recognized user's improperly-written characters.}
    \label{figure:recognized:improper:characters}
\end{figure}

An additional observation from the deployment was the types of students' improperly-written characters that were recognized by the interface.
Nine different types were noticeably observed as having consistency among students or were uniquely not expected to be written (Figure~\ref{figure:recognized:improper:characters}).

\section{Discussion and Future Work}
The results of the grading outcomes in the classroom study and the constructive feedback of students from the interaction study revealed various insights that were worth considering for improving and continuing the development efforts of \projectName{}.
The first potential next step is in regards to the visual aesthetics and the intelligent features of the interface.
Although the interaction study yielded positive feedback on the rewarding experience and intuitive interactions of the interface, students also expressed strong concerns regarding the weak visual aesthetics of the interface and occasional issues with the correctness of the animations or the discrepancies of the assessment scoring.
We suspect that these issues may stem from potential technical issues with the assessment system and the need to further tweak the values for dictating the assessment of students' character writing performance.

The second potential next step is in regards to efforts of employing previous assessment classification techniques for the interface that yielded little successes.
Specifically, one technique that was attempted was template matching for classifying the type of characters that students had written.
However, this technique was initially unsuccessful due to visual factors in how students wrote their characters that were too noisy for our implemented template-matching algorithms.
We are interested in re-visiting this approach for exploring additional assessment metrics that may be of benefit to students and of interest to instructors.

The third potential next step is in regards to further enhancing and specializing our evaluation, in order to discover additional challenges that \nth{1}- and \nth{2}-year students face in their character writing study.
Specifically, we would like to refine our classroom study's evaluation materials that incorporates more character writing-focused test materials, which we anticipate will be invaluable for more deeply revealing potential barriers---or lack of---that \nth{1}- and \nth{2}-year students encounter.
Insights from this revised evaluation will allow us to, for example, more confidently target assessments that cause issues to \nth{1}-year students and better develop assessments that better challenge \nth{2}-year students.

Overall, the successes of the interface's deployment have not only led to its continued availability to the instructors of the participating courses for subsequent semesters, but have also opened up plans to adapt this interface for characters used in Chinese foreign language classes and for character writing in more advanced Japanese foreign language classes.
We believe that \projectName{} has strong potential for becoming a richer resource supplement to foreign language courses that employ character writing.

\section{Conclusion}
In this paper, we describe our work with \projectName{}, an educational application for novice students to practice their writing in and receive intelligent assessment on their kanji script character writing.
Our work leverages various recognition techniques for assessing students' writing performance through intelligent scoring and visual animations, so that they may be better informed on their proficiency of Japanese language writing proficiency.
From our evaluations, we demonstrated that not only did students overall benefited from the interface during course enrollment, but also responded positively to the interface's various features.

\section{Acknowledgements}
We give our huge thanks to instructors George Adams and Yuki Waugh at Texas A\&M University for their valuable assistance in providing consultation on our interface and access to their students.
We also thank our fromer undergraduate students Dakota Sloniger, Rupen Sakariya, Carlo De Guzman, Jacob Mathews, and Erwin Susanto for their development work in an initial prototype of our interface.

\bibliographystyle{aaai}
\bibliography{references}

\end{document}